\documentclass[useAMS,usenatbib]{mn2e}
\usepackage{natbib}
\usepackage{epsfig}
\usepackage{amsbsy}
\usepackage{hyperref}

\providecommand{\adsurl}[1]{\href{#1}{ADS}}

\newcommand{\lya}{Lyman-$\alpha$~}

\newcommand{\mwdm}{\mbox{m$_{\rm WDM}$}}
\newcommand{\hmpc}{\mbox{$h^{-1}$\,\rm Mpc~}}
\newcommand{\ihmpc}{\mbox{$h$\,\rm Mpc$^{-1}$}}
\newcommand{\be}{\begin{equation}}
\newcommand{\ee}{\end{equation}}
\newcommand{\ba}{\begin{eqnarray}}
\newcommand{\ea}{\end{eqnarray}}
\newcommand{\brr}{\begin{array}}

\newcommand{\err}{\end{array}}
\newcommand{\bc}{\begin{center}}
\newcommand{\ec}{\end{center}}

\DeclareMathAlphabet{\mathsc}{OT1}{cmr}{m}{sc}
\def\testbx{bx}%
\DeclareRobustCommand{\ion}[2]{%
\relax\ifmmode
\ifx\testbx\f@series
{\mathbf{#1\,\mathsc{#2}}}\else
{\mathrm{#1\,\mathsc{#2}}}\fi
\else\textup{#1\,{\mdseries\textsc{#2}}}%
\fi}
\title[Non-linear Matter Power in WDM Cosmologies]
{The Non-linear Matter Power Spectrum in Warm Dark Matter Cosmologies}

\author[M. Viel et al.] 
{M.~Viel$^{1,2}$, K.~Markovi\v{c}$^{3,4,5}$, M.~Baldi$^{3,4}$, J.~Weller$^{4,5}$
\\
$^1$ INAF - Osservatorio Astronomico di Trieste, Via G.B. Tiepolo 11,
I-34131 Trieste, Italy (viel@oats.inaf.it)\\
$^2$ INFN/National Institute for Nuclear Physics, Via Valerio 2,
I-34127 Trieste, Italy\\
$^3$University Observatory Munich, Ludwig-Maximilian University, Scheinerstr. 1, 81679, Munich, Germany\\
$^4$Munich, Germany
Excellence Cluster Universe, Boltzmann Str. 2, 85748 Garching, Germany \\
$^5$Max-Planck-Institut for Extraterrestrial Physics, Giessenbachstr., 85748 Garching,
Germany\\}

\begin{document}
\maketitle
\begin{abstract}
We investigate the non-linear evolution of the matter power spectrum
by using a large set of high-resolution N-body/hydrodynamic
simulations.  The linear matter power in the initial conditions is
consistently modified to mimic the presence of warm dark matter
particles which induce a small scale cut-off in the power as compared
to standard cold dark matter scenarios. The impact of such thermal
relics is examined at small scales $k>1\,\ihmpc$, at redshifts of
$z<5$, which are particularly important for the next generation of
\lya forest, weak lensing and galaxy clustering surveys.  We
measure the mass and redshift dependence of the warm dark matter
non-linear matter power and provide a fitting formula which is
accurate at the $\sim$ 2\% level below $z=3$ and for particle masses
of \mwdm $\ge$ 0.5 keV. The role of baryonic physics on the warm dark
matter induced suppression is also quantified. In particular, we
examine the effects of cooling, star formation and feedback from
strong galactic winds. Finally, we find that a modified version
  of the halo model describes the shape of the warm dark matter
  suppressed power spectra better than {\small{HALOFIT}. In the case
  of weak lensing however, the latter works better than the former,
  since it is more accurate on the relevant, mid-range scales, albeit
  very inaccurate on the smallest scales ($k>10\,\ihmpc$) of the
  matter power spectrum.}
\end{abstract}

\begin{keywords}
Cosmology: theory -- large-scale structure of the Universe -- dark
matter, methods: numerical -- gravitational lensing: weak
\end{keywords}

\section{Introduction}

The increasing amount of observational data available and the
numerical tools developed for their interpretation have given rise
the so-called era of precision cosmology. At the present time,
the cosmological concordance model based on a mixture of cold dark
matter and a cosmological constant must thereby be tested in new
regimes (both in space and in time) and using as many observations 
and techniques as possible in order to either confirm or disprove it.

Among the many different observables the non-linear matter power
spectrum is a crucial ingredient since it allows us to describe the
clustering properties of matter at small scales and low-redshift,
where linear theory is not reliable. However, accurate modelling
of non-linear physical processes is needed, in order to use this 
observable to gain quantitative results on the nature of
dark matter.

Warm Dark Matter (WDM) is an intriguing possibility for a dark matter
candidate. Its velocity dispersions are intermediate between those of
cold dark matter and hot dark matter (e.g. light neutrinos). In this
scenario, at scales smaller than WDM free-streaming scales, cosmological
perturbations are erased and gravitational clustering is significantly
suppressed. If such particles are initially in thermal equilibrium,
they have a smaller temperature and affect smaller scales than those
affected by neutrinos. In addition, WDM produces a distinctive
suppression feature at such scales as compared to that induced by
neutrinos. For example, thermal relics of masses at around 1 keV which
constitute all of the dark matter have a free-streaming scale that is
comparable to that of galaxies, well into the non-linear regime.  Among
the different WDM candidates a special role is played by
the sterile neutrino with mass at the keV scale (\cite{boya09a}).
WDM was originally proposed to solve some putative
problems that are present in cold dark matter scenarios at small
scales (see \cite{colin00,bode}), however it is at present
controversial whether these tensions with cold dark matter predictions
can be solved by modifying the nature of dark matter particles or by
some other baryonic process (e.g. \cite{trujillo10}).

In the present paper we wish to quantify the impact of a WDM
relic on the non-linear power spectrum by using a set of
N-body/hydrodynamic simulations of cosmological volumes at high
resolution. Investigating WDM scenarios in a cosmological setting has
been done by means of N-body codes in order to carefully quantify the
impact of such a candidate in terms of halo mass function, structure
formation, halo density properties (\cite{bode,colin08,colombi09}) and
particular care needs to be placed on correctly addressing
numerical/convergence issues (\cite{ww07}). In general, while the WDM
induced suppression transfer function can be reliably estimated in the
linear regime (e.g. \cite{viel05,boya08,les11}), the non-linear suppression
has not been investigated. However, a recent attempt to obtain the non-linear
matter power at small scales by modifying the halo model is 
described in \cite{smith11}.

The analysis of matter power spectra at small scales has been
performed in recent year by different groups by focussing mostly on
baryon physics such as feedback and cooling 
(e.g.  \cite{rudd08,guillet10,casarini11,vandaalen11}).
%KM without considering how these properties are modified in WDM scenarios.  

Recently, \cite{vandaalen11} presented an extensive investigation
  of the effects of several different implementations of galactic
  feedback on the total matter power. Including feedback from Active
  Galactic Nuclei (AGN) was claimed to be most realistic since it
  matches optical and X-ray observations of groups of galaxies and
  solves the overcooling problem (see also related works by
  \cite{puchwein08,fabjan010,mccarthy10,teyssier11}). This scenario
  has a relatively large impact in terms of total matter power at the
  scales affected by the presence of a WDM candidate. In a subsequent
  paper, \cite{semboloni11} carefully analysed the impact of this
  feedback on weak lensing observables and concluded that it will not
  be possible to constrain WDM masses with future weak lensing surveys
  (as claimed in for example \cite{markovic11}). These findings are
  very interesting and show that a future measurement of the matter
  power at such small scales should be considered with an accurate
  model of baryonic physics and not only of the dark matter
  component. In the present work, although we will be exploring some
  feedback mechanism that do not include the AGN feedback, the main
  focus is on the signature of the WDM suppression in terms of total
  matter power, presented as a ratio with a corresponding $\Lambda$CDM
  standard model that includes the same astrophysical input and
  differs only in the initial total matter power.

Different constraints can be obtained by using several astrophysical
probes.  For example by using \lya observables such as the transmitted
\lya flux power, very competitive measurements in the form of lower
limits ($m_{\rm WDM}>4$ keV, 2$\sigma$ C.L.) have been derived
by using the SDSS flux power and other higher redshift and higher
resolution data \citep{viel05,viel06wdm,seljak06wdm}: these
constraints become much weaker if the WDM is assumed to account
to only a given fraction of the dark matter \citep{boya09a} or if the
initial linear suppression for a sterile neutrino is considered
\citep{boya09b} (in this latter case basically any $m_{\rm WDM,sterile}>1$ keV is
allowed). Alternatively, constraints on WDM models can be placed
using: the evolution and size of small scale structure in the local
volume high resolution simulations \citep{tikhonov09}; simulated Milky
Way haloes to probe properties of satellite galaxies
\citep{polisensky11,lovell11}; large scale structure data
\citep{abazajian06}; the formation of the first stars and galaxies in
high resolution simulations \citep{gao07}; weak lensing power spectra
and cross-spectra \citep{markovic11,semboloni11}; the dynamics of the
satellites \citep{knebe08}; the abundance of sub-structures
\citep{colin00}; the inner properties of dwarf galaxies
\citep{strigari06}; mass function in the local group as determined
from radio observations in HI \citep{zavala09}; the clustering
properties of galaxies at small scales \citep{coil08}; the properties
of satellites as inferred from semi-analytical models of galaxy
formation \citep{maccio010}; phase-space density constraints from dwarf galaxies \citep{devega10}.

We believe that most of the astrophysical probes used so far in order
to constrain the small scale properties of dark matter could benefit
from a comprehensive numerical modelling of the non-linear matter
power.  The present work aims at providing such a quantity by using
N-body/hydrodynamic simulations. The findings could also be useful
for future surveys such as PanSTARRS, DES, LSST, ADEPT, EUCLID, JDEM
or eROSITA, WFXT and SPT.

The layout of the paper is as follows.  In Section 2 we present our set
of simulations and the code we use in order to investigate the
non-linear suppression on the total matter power. Section 3 contains
the main results of the present work and the description of the checks
made in order to present a reliable estimate of the WDM non-linear
suppression: we focus on numerical convergence, box-size, baryonic
physics, particle velocities and the effect induced by cosmological
parameters on the WDM power. As an application of the findings in
Section 3 we present the weak lensing power and
cross-spectra for a realistic future weak lensing survey in Section 4 and compare
these results with those that could be obtained by using either
linear-theory or halo models in Section 5. We conclude with a summary
in Section 6.

\section{The simulations}

\begin{table}
\begin{center}
\begin{tabular}{c|c|c|}
\hline
linear size (Mpc$/h$) & \mwdm (keV) & soft. (kpc$/h$) \\
\hline
12.5 &-- & 0.62 \\
12.5 &1 & 0.62 \\
25$^{a}$   &-- & 1.25 \\
25   &1 & 1.25 \\
50   &--  &2.5 \\
50   &1  & 2.5 \\
100  &--  & 5 \\
100  &1  & 5 \\
25   &0.25 & 1.25 \\
25   &0.5 &  1.25 \\
25$^{a,b,c}$   &1   &  1.25 \\
25   &2   &  1.25 \\
25   &4   & 1.25 \\
12.5 & 1   & 0.625 \\
6.25 & 1   & 0.33 \\
\hline

\end{tabular}
\end{center}
\caption{Summary of the simulations performed. Linear box-size, mass
  of warm dark matter particle and gravitational softening are
  reported in comoving units (left, center and right columns,
  respectively).  The particle-mesh (PM) grid is chosen to be equal to
  N$_{\rm DM}^{1/3}$ with N$_{\rm DM}=512^3$. Simulations $(a)$ have been run
  with hydrodynamic processes (a simplified star formation recipe and
  radiative processes for the gas) and with full hydrodynamics with
  the standard multiphase modelling of the interstellar medium and
  strong kinetic feedback in the form of galactic winds.  Simulations
  $(a)$ have been also run at lower resolution $N_{\rm DM}=384^3$ and for
  different values of $\sigma_8\,$, $\Omega_{\rm m}$ and
  $H_0$. Simulation $(b)$ has been run by switching the initial
  velocities of warm dark matter particles off and by increasing the
  linear size of the PM grid by a factor 3. Simulation $(c)$ has been
  run with N$_{\rm DM}=640^3$ dark matter particles with a softening of 1
  kpc$/h$ to $z=0.5$.}
\label{table1}
\end{table}

Our set of simulations has been run with the parallel hydrodynamic
(TreeSPH: Tree-Smoothed Particle Hydrodynamics) code {\small
  {GADGET-2}} based on the conservative `entropy-formulation' of SPH
\citep{springel05}. Most of the runs use the TreePM (Tree-Particle Mesh)
N-body set-up and consist only of dark matter particles, however for a
few runs, in order to test the impact of baryonic physics, we switched
hydrodynamic processes on.

The cosmological reference model corresponds to a `fiducial'
$\Lambda$CDM Universe with the following parameters, at $z=0$, $\Omega_{\rm m
}=0.2711,\ \Omega_{\rm \Lambda}=0.7289,\ \Omega_{\rm b }=0.0451$,
$n_{\rm s}=0.966$, $H_0 = 70.3$ km s$^{-1}$ Mpc$^{-1}$ and
$\sigma_8=0.809$. This model is in agreement with the recent
constraints obtained by WMAP-7 year data \citep{wmap7} and by other
large scale structure probes. The initial (linear) power spectrum is
generated at $z=99$ with the publicly available software {\small{CAMB}}
{\footnote{http://camb.info/}} and then modified to simulate warm
dark matter (see below).

We consider different box sizes in order to address both the large
scale power and (more importantly) the effect of resolution.  The
gravitational softening is set to be 1/40-$th$ of the mean linear
inter-particle separation and is kept fixed in comoving units. The
dimension of the PM grid, which is used for the long-range force
computation, is chosen to be equal to the number of particles in all but
a single case in which a finer grid is used.  The simulations
follow a cosmological periodic volume filled with $512^3$ dark matter
particles (an equal number of gas particles is used for the
hydrodynamic simulations) in all but two cases in which a smaller and
a larger number of particles are chosen in order to check for numerical
convergence of matter power.  
We mainly focus on WDM
masses around 1 keV. For such a mass, the characteristic cut-off in
the power spectrum appears at scales of about $k\sim 1.5$ \ihmpc, the suppression
reaching 50\% at $k=6$ \ihmpc. The suppressed scales are highly non-linear and therefore
require high-resolution as well as N-body techniques. However,
in order to be conservative we present results for the following \mwdm
values: 0.25, 0.5, 1, 2 and 4 keV. These limits could be easily
converted to masses for a sterile neutrino particle produced in the
so-called standard Dodelson-Widrow scenario and correspond to $m_{\rm
  s}= 0.7,1.66,4.4,11.1,28.1$ keV (note that physically motivated
scenarios based on for example non-resonant production mechanisms have
been proposed, however the simulations carried out in the present work
cannot be strictly applied to those since they require a non-trivial
modification of the linear transfer function, as discussed by \cite{boya09b}).

The initial conditions for warm dark matter particles are generated
using the procedure described in \cite{viel05}, which we briefly
summarize here. The linear $\Lambda$CDM power is multiplied by the
following function:
\begin{eqnarray}
&&T^2_{\rm lin}(k)\equiv P_{\rm WDM}(k)/P_{\rm \Lambda CDM}(k)=(1+(\alpha\,k)^{2\nu })^{-5/\nu}
\nonumber , \\ && \alpha(\mwdm)=0.049\,\left(\frac{1
  \rm{keV}}{\mwdm}\right)^{1.11}\,\left(\frac{\Omega_{\rm WDM}}{0.25}\right)^{0.11}\left(\frac{h}{0.7}\right)^{1.22}
\label{eq1}
\end{eqnarray}
where $\nu=1.12$ and $\alpha$ has units of $h^{-1}$ Mpc
\citep[e.g.][]{hansen02}.  We stress that the above equation is an
approximation which is strictly valid only at $k<5-10$ \ihmpc. Below this
scale the warm dark matter power spectrum could be described by a more
complicated function and acoustic oscillations are present \citep[see for
example the recent work in][]{les11}.
 
Initial velocities for warm dark matter
particles are drawn from a Fermi-Dirac distribution and added to the
proper velocity assigned by linear theory: the r.m.s. velocity
dispersion associated to their thermal motion is 27.9, 11.5, 4.4. 1.7,
0.7 km/s for \mwdm=0.25,0.5,1,2,4 keV, respectively. The typical
r.m.s. velocity dispersion for the dark matter particles of the
$\Lambda$CDM runs is $\sim 27$ km/s, so at least for masses above 1
keV the thermal WDM motion is a small fraction of the physical
velocity dispersion assigned by the Zel'dovich approximation.

\begin{figure*}
%[H]
\begin{center}
\includegraphics[width=14cm]{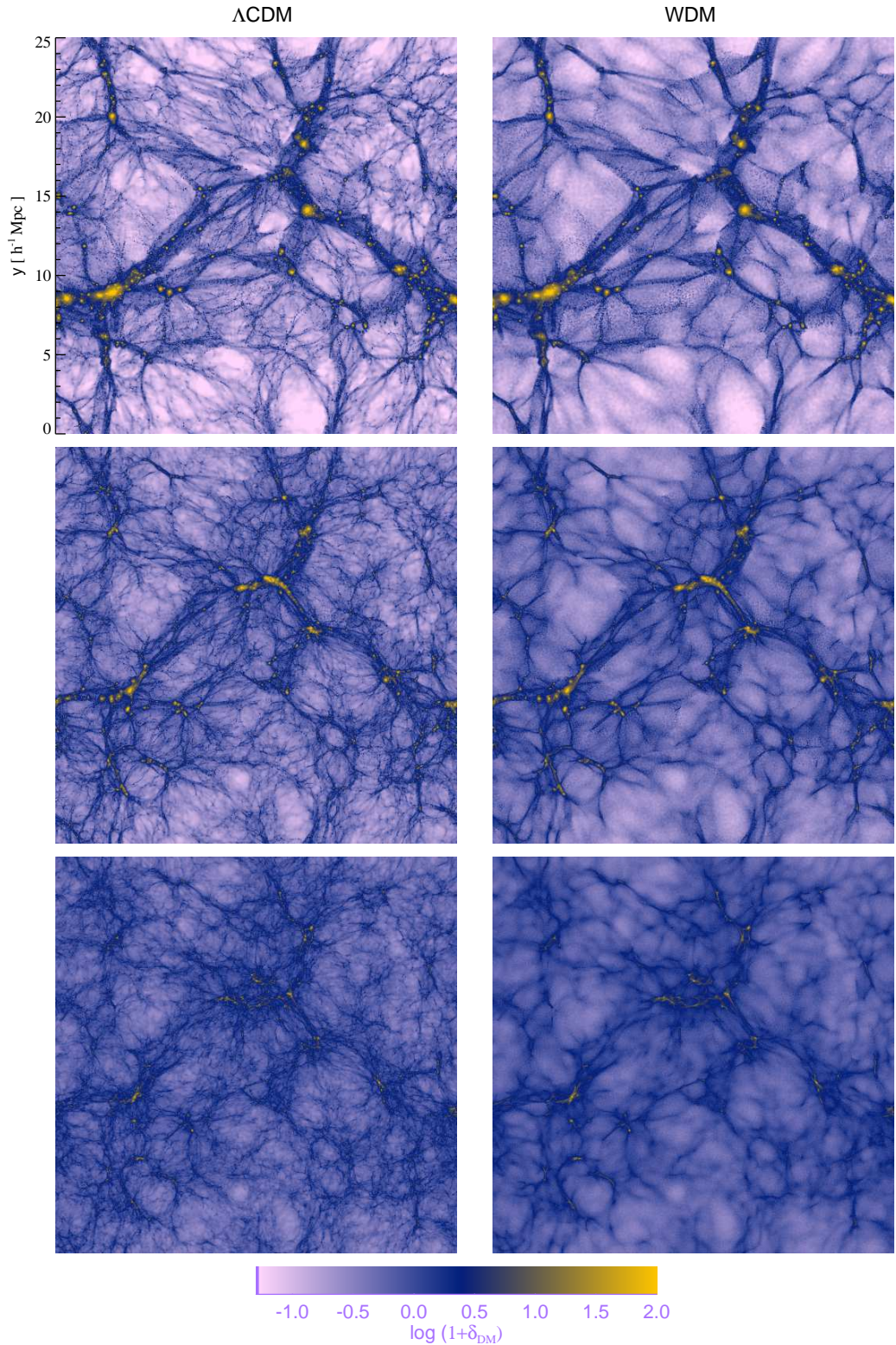}
\end{center}
\caption{``Visual'' inspection of the redshift evolution of cosmic
  structures in the $\Lambda$CDM and WDM (\mwdm=1 keV) scenarios (left
  and right columns, respectively) for the defaults (25,512)
  runs. From the top to the bottom rows we show a 2.5 \hmpc thick
  slice of the projected dark matter density at $z=0,2,5$
  respectively. At $z=0$ the clustering properties of the dark matter
  at scales $k<10$ \ihmpc are indistinguishable in the two scenarios,
  while at $z=2,5$ the WDM model has a suppression in power of about 5\%
  and 25\% at $k=10$ \ihmpc.}
\label{fig_slice}
\end{figure*}

When baryonic physics is included, we consider the following processes:
$i)$ radiative cooling and heating, $ii)$ star formation processes,
$iii)$ feedback by galactic winds. We note that metal cooling is
  not included in the simulations and only cooling from H and He is
  considered, while galactic winds are powered by massive stars and
  not AGN.

The rationale is to see at which level these processes
impact the non-linear matter power at small scales in order to compare to the suppression effects of 
WDM. Thus, we are not aiming at exploring in a
comprehensive way the impact of these processes on the non-linear
power at small scales.  \citep[e.g.][]{vandaalen11,casarini11}: the
baryonic simulations are used only to quantify the impact of such
processes on the suppression induced by WDM w.r.t. cold
dark matter scenarios.

Radiative cooling (H and He) as well as heating processes are assumed for
a primordial mix of hydrogen and helium corresponding to a mean Ultraviolet
Background similar to that produced by quasars and galaxies and
implemented in \cite{katz96}. This background naturally gives a
hydrogen ionization rate $\Gamma_{-12}\sim 1$ at high redshift and an
evolution of the physical state of the intergalactic medium (IGM)
which is in agreement with observations \citep[e.g.][]{bolt05}.  The
star formation criterion for the default runs is a very simple one
that converts all the gas particles whose
temperature falls below $10^5$ K and whose density contrast is larger
than 1000 into collisionless stars \citep[more details can be found in][]{viel04}. This
prescription is usually called ``QLYA'' (quick \lya) since it is very
efficient in quantitatively describing the \lya forest and the low
density IGM. We also run a simulation with the full multi-phase
description of the interstellar medium (ISM) and with kinetic feedback
in the form of strong galactic winds as in \cite{Springel03}. The
chosen speed of the wind is 483 km/s and both the ISM modelling and
this feedback mechanism is expected to impact on the distribution of
baryons and thus on the total matter power spectrum. The wind
  particles temporarily decouple from the hydrodynamics: the
  maximum allowed time of the decoupling is $t_{\rm dec}= l/v_{\rm w}$
  with l=20 kpc$/h$ and $v_{\rm W}=483$ km/s (see for example
  \cite{dallavecchia08} and \cite{pierleoni08} to understand the
  effects of this parameter in terms of feedback efficiency and on
  the properties of local HI-galaxies).

We note that simulations that include baryons are significantly slower
than the default dark matter only runs and therefore our constraints
will mainly be derived from the former simulations.

In the
following, the different simulations will be denoted by two numbers,
$(N_1,N_2)$: $N_1$ is the size of the box in comoving Mpc$/h$ and
$N_2$ is the cubic root of the total number of gas particles in the
simulation.
The mass per dark matter particle is $8.7\times10^6$M$_{\odot}/h$ for
the default (25,512) simulations. This mass resolution allows to adequately
sample the free-streaming mass for the models considered here.  

In Figure \ref{fig_slice} we show the projected dark matter density as
extracted from the default (25,512) runs in the $\Lambda$CDM case
(left) and WDM case (right) for \mwdm=1 keV. This WDM particle mass is
already ruled out at a significant level by \lya forest observations
\citep[e.g.][]{seljak06,viel06}.  The different rows refer to
$z=0,2,5$ from top to bottom, respectively. In this Figure it is
essential to see how the clustering proceeds differently in the two
scenarios and while there are large differences below the Mpc scale at
$z=5$ between the two cosmic webs, these differences are largely
erased by non-linear evolution at $z=0,2$.

The main features of the simulations are summarized in Table 1.

\begin{figure*}
\begin{center}
\includegraphics[width=14cm]{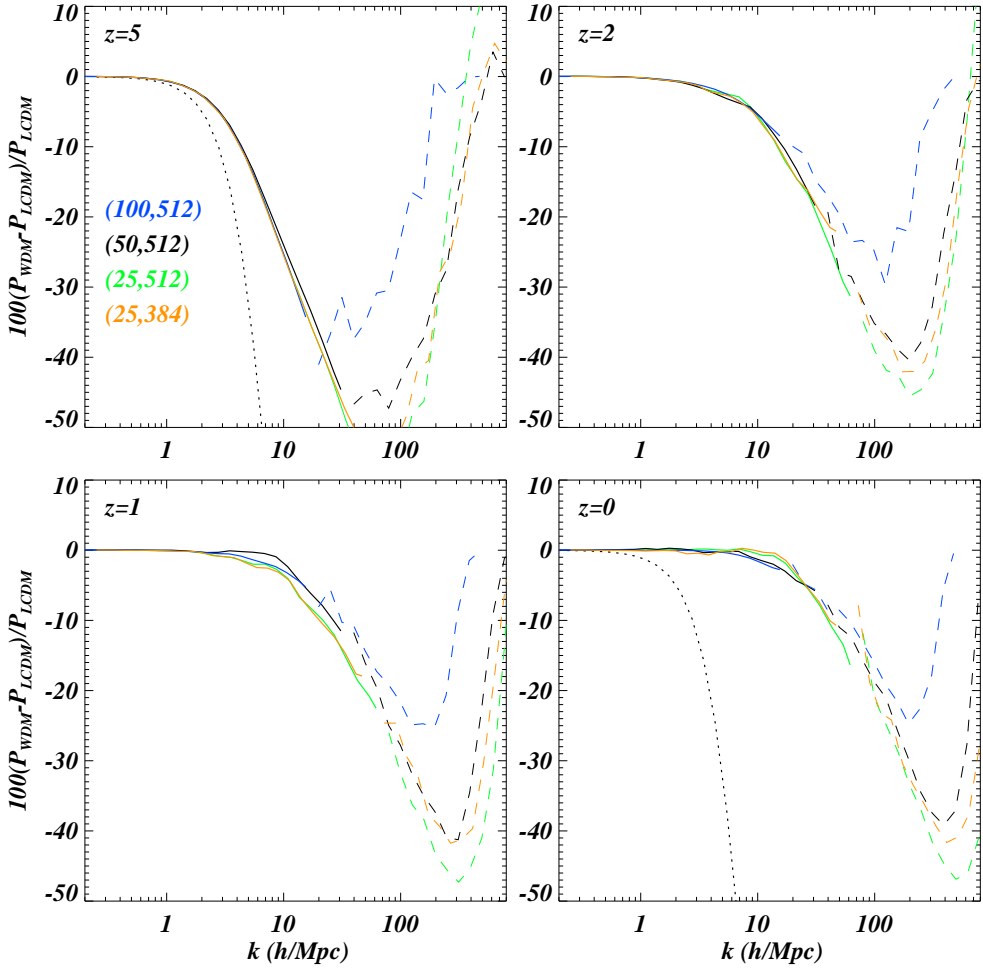}
\end{center}
\caption{Percentage difference between warm dark matter non-linear
  power and cold dark matter for the different runs. The mass of the
  warm dark matter particle is kept fixed to $\mwdm=1$ keV. Blue,
  black, green curves refer to 100, 50, 25 \hmpc respectively and with
  a fixed number of particles $N_{\rm DM}=512^3$. The orange curves
  refer to 25 \hmpc and has a fixed number of particles $N_{\rm
    DM}=384^3$ The continuous lines represent the large scale estimate
  of the power, while the dashed ones describe the small scale power
  obtained with the folding method (see text). The four panels
  represent different redshifts at $z=0,1,2,5$ (bottom right, bottom
  left, top right and top left, respectively). The dotted line plotted
  at $z=0$ and $z=5$ is the redshift independent linear suppression
  between the two models.}
\label{fig_res}
\end{figure*}
\begin{figure*}
\begin{center}
\includegraphics[width=14cm]{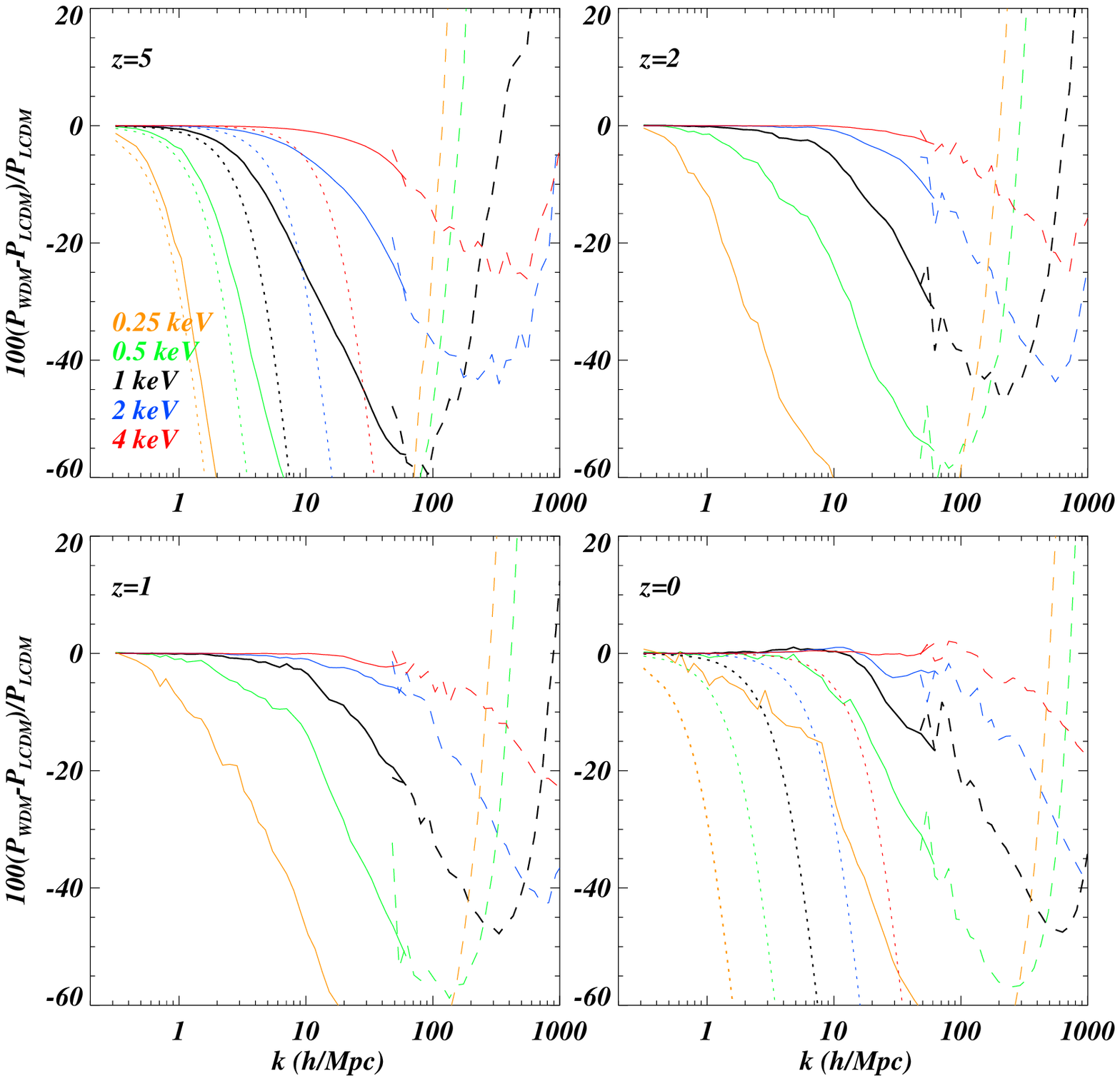}
\end{center}
\caption{Percentage difference between warm dark matter non-linear
  power and cold dark matter for the different runs. The resolution is
  kept fixed in this plot and only 25 \hmpc boxes are
  considered. Orange, green, black, blue and red curves refer to
  $\mwdm=0.25,0.5,1,2,4 $ keV, respectively. The continuous lines represent
  the large scale estimate of the power, while the dashed ones
  describe the small scale power obtained with the folding method (see
  text). The four panels represent different redshifts at $z=0,1,2,5$
  (bottom right, bottom left, top right and top left,
  respectively). The dotted coloured curves plotted at $z=0$ and $z=5$
  are the redshift independent linear suppression between the
  different models.}
\label{fig_mass}
\end{figure*}

\section{Results}
In this section we describe the main results obtained from our sample
of simulations. The power is computed from the distributions of the
different sets of particles (dark matter, gas and stars) separately
and for the total matter component by performing a CIC (Cloud-In-Cell)
assignment to a grid of the size of the PM grid. The CIC kernel
is also deconvolved when getting the density at the grid points
\citep[e.g.][]{vhs010}). We also show a small scale estimate ($k > 10$
\ihmpc) of the power obtained with the folding method described in
\citep{jenkins98,colombi09}, although this power will not be used
quantitatively.

We will plot the suppression in power as a percentage difference
between WDM and $\Lambda$CDM total matter power spectra, normalized by
the default $\Lambda$CDM total matter power. The initial conditions
for CDM and WDM have the same phases and cosmological/astrophysical
parameters in order to highlight the effect of the warm dark matter
free streaming.
 
\subsection{Resolution and box-size}
In Figure~\ref{fig_res} we show the percentage difference between the
total non-linear powers of WDM (\mwdm=1 keV) and $ \Lambda$CDM
runs. We subtract the shot-noise power from all the power spectrum
estimates made. For our largest box-sizes the shot-noise
power is comparable to the actual measured power at $z=0$ at $k\sim
150$ \ihmpc, while for the default simulations (25,512) of \mwdm=1
(0.25) keV the matter power is always above the shot-noise level for
$z<10$ and for $k<20 (7)$ \ihmpc.

This figure focusses on the resolution and box-size effects and presents
the percentage difference at four different redshifts $z=0,1,3,5$ (bottom right, bottom
left, top right and top left panels respectively) and for three
different box-sizes (100, 50, 25 \hmpc\,  shown as blue, black and green
curves respectively). The dotted line represents the redshift independent linear
cut-off of Eq.\ref{eq1}, while the lower resolution (25,384) run is
also plotted in orange. Here, there are two estimates for the power:
one at large scales (continuous curves), the second at smaller scales
(dashed curves). We are primarily interested in the power at scales
$k<10$ \ihmpc and therefore only the large scale estimate will be used.
However, we also show the power at smaller scales since
physical and numerical effects play a larger role in this range. We
note that the linear theory suppression is a good approximation only
at $k<1$ \ihmpc.  From the figure one can see that there is
convergence up to $k=50$ \ihmpc between (25,512) and (25,384) runs in
the redshift range considered. The resolution used is thus
sufficient for \mwdm=1 keV particles. Note that
\citet{vandaalen11} recently found that (100,512) $\Lambda$CDM
simulations have sufficiently converged at scales $k<10$ \ihmpc.  At
$k=3 (10)$ \hmpc ~and $z=5$ there is already a 5 (50)\% difference
between the linear and non-linear power. At $z=0,1,3$ the differences
between WDM and $\Lambda$CDM power is below 1\%, 2\% and 5\% respectively at
$k=10$ \ihmpc.  The maximum suppression dip is strongly influenced by
resolution and moves to larger wavenumbers when the resolution
increases.  At $k>100$ \ihmpc we note a steep (resolution dependent)
turn-over in the suppression which is likely to be due to effects that
impact on the halo structure and which has also been found in CDM
numerical simulations that include a fraction of the matter content in
the form of active neutrinos \citep{brandbyge08,vhs010}.

We have checked that increasing the particle-mesh grid by a factor
three (i.e. PM=1536) has negligible impact on the total matter power
at scales $k<100$ \ihmpc.  In order to test the robustness of our
results in terms of shot-noise level we have also run a WDM simulation
with \mwdm=1 keV and $N_{\rm DM}=640^3$ particles and compared the
power spectra with the (25,512) and (25,384) runs: we confirm very
good agreement between these simulations at $k<20$ \ihmpc in the
redshift range considered in the present work. More precisely, the
(25,512) and (25,640) WDM runs agree below the one percent level at
$k<100$ \ihmpc.

\subsection{The effect of the mass of a warm dark matter particle}

Here we address the effect of varying \mwdm ~on the
non-linear matter power.  The results are shown in Figure
\ref{fig_mass} where we report five different masses for the (25,512)
default runs.  The curves correspond to \mwdm=0.25,0.5,1,2 and 4 keV
(orange, green, black, blue and red curves, respectively) at $z=0,1,2$
and $5$ (bottom right, bottom left, top right and top left,
respectively). The linear suppressions are also shown with dotted
lines of the corresponding colors.  At $z=5$ we can see large
differences between the models that become smaller with the redshift
evolution.  The 20\% suppression at $k=10$ \ihmpc at $z=5$ for the \mwdm=1 keV
model becomes 2\% at $z=1$ and it is below 1\%  at $z=0$: the
clustering properties of the dark matter are the same at scales above
$k\sim 10$ \ihmpc at least for \mwdm$>1$ keV. The \mwdm=0.5 keV model
still presents a 7\% suppression by $z=0$, while the suppression is
four times larger at $z=2$.  The linear suppression is a very poor
approximation in the range of wavenumbers considered here even at high
redshift.  At $z=1$, which is particularly interesting for weak
lensing data, a 2\% measurement of the non-linear power is likely
to be able to exclude models below the 1 keV value (bottom left
panel).  The dip of the maximum suppression and the turn-over both move to
larger scales as the mass decreases.

We have also investigated the importance of WDM velocities
in the initial conditions by running a simulation without assigning a 
Fermi-Dirac drawn thermal velocity to the dark matter particles.
We tested this for a \mwdm=1 keV model and found differences always 
below 1\% in terms of total matter power at the scales of interest here.

\subsection{Baryonic effects}

\begin{figure*}
\begin{center}
\includegraphics[width=14cm]{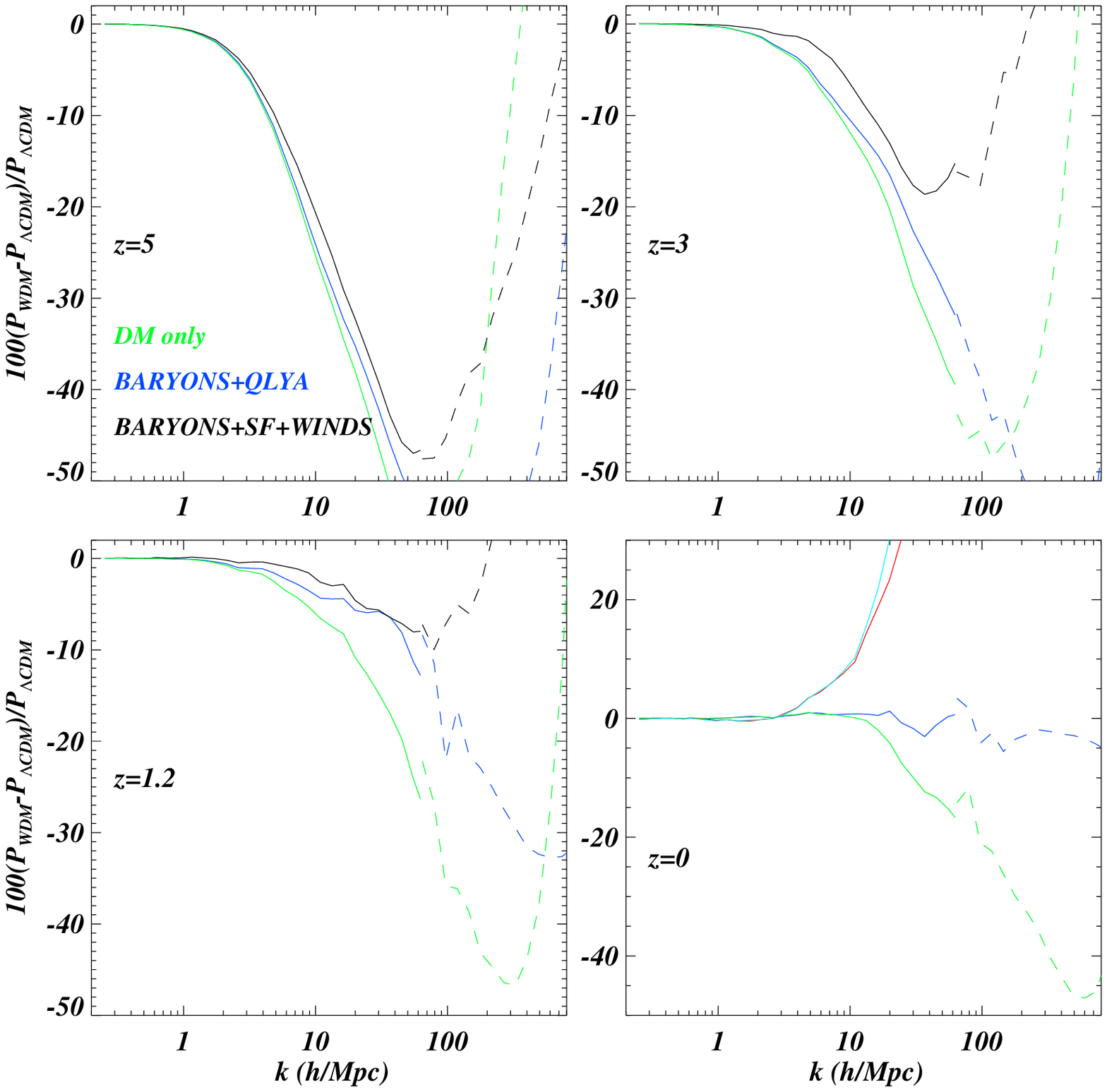}
\end{center}
\caption{Percentage difference between total WDM non-linear
    matter power and total $\Lambda$CDM non-linear matter power for
    runs that incorporate baryonic physical processes. The
  simulations refer to a 25 \hmpc box and $m_{\rm WDM}$=1 keV. The
  green curves refer to the pure dark matter simulations; blue curves
  refer to simulations that include baryons, cooling from H and He and
  a simplified recipe for star formation that turns into collisionless
  stars all the gas particle below T=10$^5$ K and denser than 1000
  times the mean density (QLYA); black curves are instead obtained by
  using the default criterion of multi-phase star formation of
  Springel (2005) and feedback in the form of strong kinetic driven
  winds (this simulation was stopped at $z=1.2$). The continuous lines
  represent the large scale estimate of the power, while the dashed
  ones describe the small scale power obtained with the folding method
  (see text). The four panels represent different redshifts at
  $z=0,1.2,3,5$ (bottom right, bottom left, top right and top left,
  respectively). In the $z=0$ panel (note the different scale for the
  $y-$axis) we also show as the red and cyan curves the percentage
  difference between total matter power spectra that include and
    do not include cooling for $\Lambda$CDM (red) and WDM (cyan)
    models: namely the quantity $100\times (P_{\rm mat}^{\rm
      baryons+QLYA}-P_{\rm mat}^{\rm DMONLY})/P_{\rm mat}^{\rm
      DMONLY}$.}
\label{fig_baryons}
\end{figure*}

In this section we explore the effects of baryonic physics on the warm
dark matter suppression.  Baryons amount to about 17\% of the total
matter content and we expect that astrophysical processes affecting
their properties can impact the total matter power at small scales
at some level. We identify three processes that are able to modify the
clustering properties of baryons: radiative processes, star formation
and galactic feedback.  These processes are usually modelled by
hydrodynamic simulations of galaxy formation. Here, the main goal is
not to explore fully the many parameters governing these important
physical aspects, but rather to address their impact in 
WDM models by adopting prescriptions that are widely used in
the literature. There could well be other astrophysical processes
(radiative transfer effects, feedback from active galactic nuclei,
etc.) that can also affect the distribution of baryons and their
clustering properties \citep[see for example][]{vandaalen11}.

In Figure \ref{fig_baryons} we plot the WDM suppression
for the default simulation of \mwdm=1 keV for three different cases:
pure dark matter (green curves); a hydrodynamic simulation that
includes cooling (H and He only) and heating by an ultraviolet background as well as the
simple star formation criterion able to simulate the \lya forest
(``BARYONS+QLYA'' run in blue); a hydrodynamic simulation that
includes the full star formation model
based on the multi-phase description of the ISM (sophisticated compared to QLYA)
and strong galactic feedback in the form of
winds. (``BARYONS+SF+WINDS'' in black).  Unfortunately, due to the
fact that hydrodynamic simulations are slower than dark matter only
runs it was not possible to carry this last simulation down to
$z=0$ and it was stopped at $z=1.2$.

We also report the ratio between mass in stars $\Omega_{\rm
    WDM}^{*}/\Omega_{\rm \Lambda CDM}^{*}(z=0,1,2,3) = 0.85,0.78, 0.7,
  0.6$ for the ``BARYONS+QLYA'' runs, while we have $\Omega_{\rm
    WDM}^{*}/\Omega_{\rm \Lambda CDM}^{*}(z=1.2,3) = 0.87, 0.63$ for
  the ``BARYONS+SF+WINDS''runs with galactic winds feedback. For both
  the WDM and $\Lambda$CDM runs the mass fraction in stars is reduced
  by a factor five when winds are included compared to the
  ``BARYONS+QLYA'' case.

All of these processes can significantly change the clustering of
baryons especially at intermediate scales where baryon pressure is
important ($k\sim 1$ \ihmpc), where they are not expected to trace the
dark matter and at smaller scales due to the complex interplay between
feedback and star formation processes. Cooling as well as heating
modify the thermal properties of the gas and are important especially
for the low density IGM; the star formation criterion determines how
much gas is turned into stars within the potential wells of dark
matter haloes; galactic winds displace gas out of the galaxies into
the low density IGM, usually in a hot phase that prevents subsequent
cooling. Since the cosmic structure is generally different in CDM and
WDM models we do not expect the WDM suppression to be exactly the same
between two simulations that share the same astrophysical
prescriptions.  From Figure \ref{fig_baryons}, one can see that dark
matter only simulations are in good agreement (at the percent level up
to $k=10$ \ihmpc) with simulations that include radiative cooling
(metal cooling is not included) and QLYA star formation, while at
smaller scales there are significant differences. It is clear that the
presence of baryons and star formation greatly affects the maximum
suppression and the turn-over. Note that differences much larger
  than 10\% between simulations implementing different radiative
  processes (e.g. metal cooling) or feedback recipes are expected at
  $k>20$ \ihmpc in $\Lambda$CDM models \citep[see
    e.g.][]{rudd08,guillet10,vandaalen11}. Furthermore, in the case of
  AGN feedback at the level required to match observed gas fractions
  of groups a 10\% difference is found already at $k=1$ \ihmpc and a
  1\% reduction already at $k=0.3$ \ihmpc (\cite{vandaalen11}).
  
  In the $z=0$ panel we also show the difference in the power spectra
  of $\Lambda$CDM and WDM models by normalizing to the corresponding
  dark matter only model, in order to highlight the effect of cooling
  produced by baryons as opposed to the WDM signature. The two
  percentage differences are shown as cyan (WDM) and red
  ($\Lambda$CDM) curves: the WDM universe when filled with baryons
  that can cool has more power than a corresponding $\Lambda$CDM
  universe filled with the same baryon fraction. The quantity $P_{\rm
    nl,WDM,cooling}/P_{\rm nl,WDM,dmonly}$ is about 5\% larger than
  $P_{\rm nl,\Lambda CDM,cooling}/P_{\rm nl,\Lambda CDM,dmonly}$ at
  $k=10$ \ihmpc and $z=5$, at $z=1$ it becomes only 2 \% larger and by
  $z=0$, there are no differences between the two quantities at $k=10$
  \ihmpc.  The cooling of baryons inside the potential wells of dark
  matter haloes produces further collapse of structures and in general
  increases the (total) matter power spectrum.  It is thus likely than
  in the WDM model the baryons cool slightly more efficiently than in
  the corresponding $\Lambda$CDM since at high redshifts, the collapse
  of haloes around the WDM cutoff is rapid and small scale modes
  affected by cooling (H and He) grow more rapidly than in CDM: this
  is also the trend found by \cite{gao07} from the analysis of cooling
  at very high resolution and high redshift in hydrodynamic
  simulations.
\begin{figure*}
\begin{center}
\includegraphics[width=14cm]{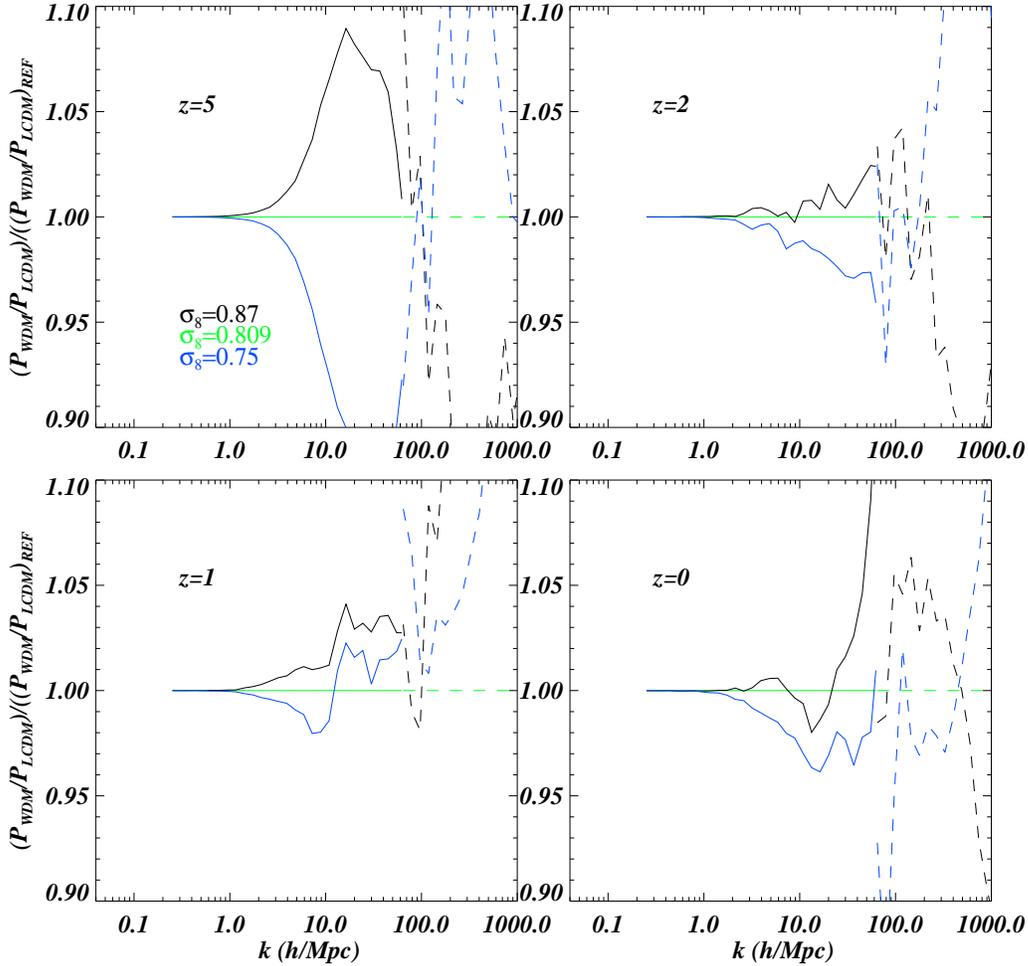}
\end{center}
\caption{Impact of a different $\sigma_8$ value in terms of
  WDM-induced suppression. The four panels represent different
  redshifts at $z=0,1,2,5$ (bottom right, bottom left, top right and
  top left, respectively) for the (25,512) with \mwdm=1 keV. Green
  represents the ($\sigma_8=0.809$) reference case, while the two
  other curves indicate the suppression for $\sigma_8=0.87$ (black)
  and $\sigma_8=0.75$ (blue).}
\label{sig8diff}
\end{figure*}

The WDM suppression is thereby highly influenced by astrophysical
effects at $k=100$ \ihmpc. In general we expect an additional
suppression due to baryons of about 2-3\% at $k=10$ \ihmpc at $z>1.5$
for \mwdm=1 keV, while this discrepancy becomes smaller at lower
redshifts. The numbers above do not apply once AGN feedback is
  included and are greatly underestimated, if AGN feedback impacts the
  matter power at the level found by \cite{vandaalen11} and
  \cite{semboloni11}.  In order to accurately measure power on these
  scales, any such AGN feedback should be accounted for.

\subsection{Other cosmological parameters}
To test the robustness of our results we extended the set of
simulations by exploring also other cosmological parameters, namely:
$\Omega_{\rm m}$, H$_0$ and $\sigma_8$.  In order to do that we modify
the input linear $\Lambda$CDM parameter calculated by {\small{CAMB}}
and vary one parameter at a time. It is clear that some parameters
like $\sigma_8$ (or $A_s$) do not have any impact at the linear level,
while they could impact the non-linear power in a way that should by
quantified with simulations.  We choose the following parameters for
the WDM and corresponding $\Lambda$CDM runs: $\Omega_{\rm
  m}=0.22,0.32$, $H_0=62,78$ km/s/Mpc and $\sigma_8=0.75,0.87$. When
calculating the suppression we always normalize both simulations to
the same $\sigma_8$ value ($\sigma_8=0.75, 0.809$ and $0.87$).  Since the
WDM suppression of the power spectrum has a relatively distinct shape
and the cut-off scale is at much higher $k$ than what is probed by
$\sigma_{8}$ normalisation, we expect WDM to be nearly independent of
any other parameter probed. The range explored by the $H_0$ values
produces a maximum $\pm 2$\% difference in terms of the WDM
suppression compared to the reference $H_0=70.3$ km/s/Mpc case at
$k=1-10$ \ihmpc and at $z<3$, while at $z=5$ there is a 5\% difference
at $k=10$ \ihmpc. The $\Omega_{\rm m}$ parameter produces a maximum
difference of 1\% at $z<3$ in the same range of wavenumbers and about
5\% at $z=5$ and $k=10$ \ihmpc. A different choice of $\sigma_8$ has a
slightly larger impact. This is seen in Figure \ref{sig8diff}), where
the WDM induced suppression with such a choice of $\sigma_8$ is
divided by the reference case of $\sigma_8=0.809$. It is clear from
the figure that the large (10 \%) differences in place at $z=5$ are
largely canceled by the non-linear growth and are at the $\pm 2$\%
level at $z=1-2$ and at the 3\% level at $k=10$ \ihmpc today.

Motivated by the present findings we regard our non-linear cutoff and
its redshift dependence as robust at least for the range of
cosmological parameters investigated at $z<3$, for \mwdm $\ge$ 0.5 keV
and at $k= 1-10$ \ihmpc: in fact the differences are at the $\pm$ 2\%
level and in the next section we will provide a fitting formula with a
comparable level of accuracy.  Larger masses for \mwdm ~will only
result in smaller differences in terms of WDM suppression.

We also notice that degenerate features with the nonlinear WDM
suppression might arise in the context of non-standard models of dark
energy, as e.g. interacting dark energy scenarios \citep[see
  e.g.][]{Baldi_2010b}. The investigation of such possible
degeneracies goes beyond the scope of the present paper.

\subsection{An analytical fitting formula}
Inspired by the corresponding formula for the linear suppression, we
have found the following fitting formula to be a good approximation of
the late time evolution of the non-linear suppression with an accuracy at
the 2\% level at $z<3$ and for masses larger than \mwdm=0.5 keV:

\begin{eqnarray} \label{eq_fitting}
&&T^2_{\rm nl}(k)\equiv P_{\rm WDM}(k)/P_{\rm \Lambda CDM}(k)=(1+(\alpha\,k)^{\nu l})^{-s/\nu} \nonumber , \\
&& \alpha(\mwdm,z)=0.0476\,\left(\frac{1 \rm{keV}}{\mwdm}\right)^{1.85}\,\left(\frac{1+z}{2}\right)^{1.3} \, ,
\end{eqnarray}
with $\nu=3$, $l=0.6$ and $s=0.4$.

We have chosen as a pivot redshift $z=1$ since this is the redshift
where accurate weak lensing data will be available.
This formula has been derived from the dark matter only runs.

%%%%%%%%%%%%%%%%%%%%%%%%%%%%%%%%%
\begin{figure*}
\begin{center}
\includegraphics[width=18cm]{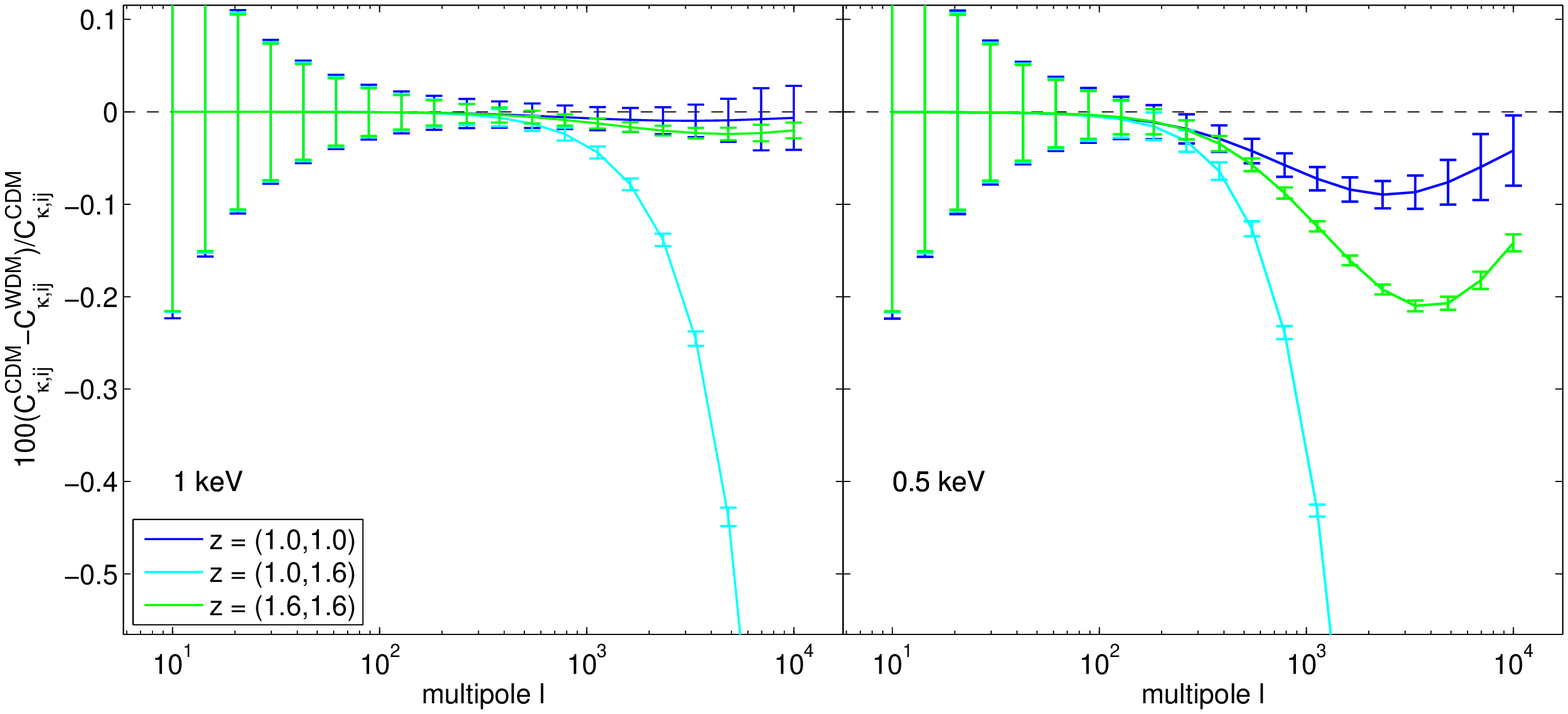}
\end{center}
\caption{The percentage WDM effect in auto- and cross-correlation
  power spectra of redshift bins at approximately $z=1$ and $z=1.6$,
  respectively. All the lines are calculated from non-linear matter
  power spectra modified for WDM by the fitting function in Equation
  \ref{eq_fitting} for WDM particle masses of 1 keV (left panel) and
  0.5 keV (right panel). In addition we plot predicted error bars for
  a future weak lensing survey, dividing the multipoles into 20
  redshift bins. Note that the error bars on auto and cross power
  spectra of different bins are correlated and therefore in order to
  fully characterise the detectable differences between the WDM (solid
  lines) and CDM (dashed black line at 0) models, one must know the
  entire covariance matrix for a survey. Note secondly that the auto
  power spectra of redshift bins at $z=1$ and $z=1.6$ have an upturn
  around $l \sim 10^3$. This is due to the dominance of shot noise on
  those scales. This upturn is not present in the cross power
  spectrum, because through cross correlation this noise due to
  intrinsic galaxy ellipticities is eliminated.}
\label{fig_Clbins}
\end{figure*}

\section{Weak lensing shear power spectra} \label{sec_wlps}
Following \cite{markovic11} and \cite{smith11}, we examine the effect of 
the fitting function in Equation \ref{eq_fitting} on the weak lensing power 
spectrum. Weak gravitational lensing is the distortion found in images of 
distant galaxies due to the deflection of light from these galaxies by the 
gravitational potential wells of intervening matter. For a review, see for 
example \cite{bartelmann2001}. The advantage of gravitational lensing 
is that unlike other large scale structure data, it does not require a 
knowledge of galaxy bias for the derivation of the properties of the 
underlying dark matter density field and is, at least on large scales, 
independent of baryonic physics. In other words, the weak lensing power 
spectrum directly probes the matter power spectrum. However, weak 
lensing measures the matter power spectrum at low redshifts. For this 
reason it is necessary to have available robust models of non-linear 
structure. For a survey able to probe angular multipoles from $l\sim20$ 
up to $l\sim 2\times10^4$, in the redshift range of $z=0.5-2.0$, the 
corresponding range of wavenumbers must be $k\sim 0.005-15$~\ihmpc. 
Note that the matter power at $k>10$~\ihmpc only has a significant 
contribution to the weak lensing power spectrum at lower redshifts, where 
however the lensing power is lower.

Future weak lensing surveys accompanied by extensive photometric 
redshift surveys will be able to disentangle the contribution to weak lensing 
by dark matter at different redshifts, by binning source galaxies into 
tomographic bins \citep{hu1999}. By cross and auto correlating  the 
lensing power in these bins, the three dimensional dark matter distribution 
can be reconstructed. An existing example of such a reconstruction is the 
COSMOS field \citep{massey2007}. Such tomography probes the non-linear matter 
power spectrum at different redshifts.

We use {\small{HALOFIT}} \citep{smith} to calculate non-linear corrections to 
the approximate linear matter power spectrum \citep{ma1996}. We then 
apply Equation \ref{eq_fitting} to approximate the WDM effects and find 
the weak lensing power spectrum \citep[e.g.][]{takada2004}:
\begin{equation}\label{eq_lensing}
C_{ij}(l) = \int_{0}^{\chi_{\rm H}} d\chi_{\rm l} W_{i}(\chi_{\rm l})W_{j}(\chi_{\rm l})\chi_{\rm l}^{-2}P_{\rm nl}\left(k=\frac{l}{\chi_{\rm l}},\chi_{\rm l}\right) \, ,
\end{equation}
where $\chi_{\rm l}(z_{\rm l})$ is the comoving distance to the lens at 
redshift $z_{\rm l}$ and $W_{i}$ is the lensing weight in the tomographic 
bin {\it i}:
\begin{equation}
W_{i}(z_{\rm l}) = \frac{4\pi G}{a_{\rm l}(z_{\rm l})c^2}\rho_{\rm m,0} \chi_{\rm l}\int_{z_{\rm l}}^{z_{\rm max}} n_{i}(z_{\rm s})\frac{\chi_{\rm ls}(z_{\rm s},z_{\rm l})}{\chi_{\rm s}(z_{\rm s})} dz_{\rm s}  \hspace{0.3cm} ,
\end{equation}
where we assume a flat universe and $a_{\rm l}(z_{\rm l})$ is the scale 
factor at the redshift of the lens, $\rho_{\rm m,0}$ is the matter energy 
density today and $n_{i}(z_{\rm s})$ is the normalised redshift distribution 
of sources in the $i$-th tomographic bin. We bin the multipoles into 20 bins.

In order to assess detectability of WDM by future weak lensing surveys,
we calculate predicted error bars on the weak lensing power spectrum
using the covariance matrix formalism \citep[][]{takada2004} and assuming errors for a
future realistic weak lensing survey as in \cite{markovic11} and
\cite{smith11} with 8 redshift bins in the range $z=0.5-2.0$. We plot
the resulting percentage differences between WDM and CDM weak lensing
power spectra in Figure \ref{fig_Clbins}. It is important to note that
the error bars in the figure do not fully characterise the sensitivity
of the power spectra, since there are additional correlations between
the error bars of different bin combinations. Additionally, there are
correlations in the error bars on large $l$ (small scales) due to
non-linearities. Further statistical tests using the entire covariance
matrix must be used in order to fully account for the above
correlations. For this plot we choose only the 5$-th$ and 8$-th$
redshift bins, whose source galaxy distributions have the mean at
$z\sim1.0$ and $1.6$ respectively. These bins are chosen because they
represent a range with the maximal WDM effect as well as lensing
signal. Note that the upturn around $l \sim 10^3$ in the
auto-correlation power spectra of bins 5 and 8 is due to the dominance
of shot noise on those scales. This noise is due to intrinsic galaxy
ellipticities and can be eliminated by cross-correlating different
redshift bins, as can also be seen in Figure \ref{fig_Clbins} \citep[see
also][]{takada2004}.

In the right panel of Figure \ref{fig_Clbins} we plot the effects of
the $0.5$ keV particle and since the black dashed line lies far
outside the error bars this is a strong indication that such a
particle can be ruled out (or detected) by a future weak lensing
survey.  This is consistent with previous works
\citep{markovic11,smith11}. In the left panel of Figure
\ref{fig_Clbins} we plot the effects of a $1$ keV WDM particle: in
this case it is more difficult to distinguish from CDM (black dashed
line), but the strongly affected cross power spectra are still
significantly different from their expected values in $\Lambda$CDM.

In a recent paper, \cite{semboloni11} have explored the effect
that AGN feedback has in terms of matter power and weak lensing power spectra
finding that there is a suppression of about 30\% at $k=10$ \ihmpc
when this feedback mechanism is included. Although they did not
investigate WDM models, this result is important, since it shows that
the effect could be much larger than the corresponding WDM induced
suppression and comparable at $z=0$ to the \mwdm=0.25 keV case.  It is
clear that future weak lensing surveys aiming at measuring the matter power at
these scales should carefully consider AGN effects since they could be
degenerate with cosmological parameters such as the mass of the
WDM particle.

\begin{figure*} 
\begin{center}
\includegraphics[width=18cm]{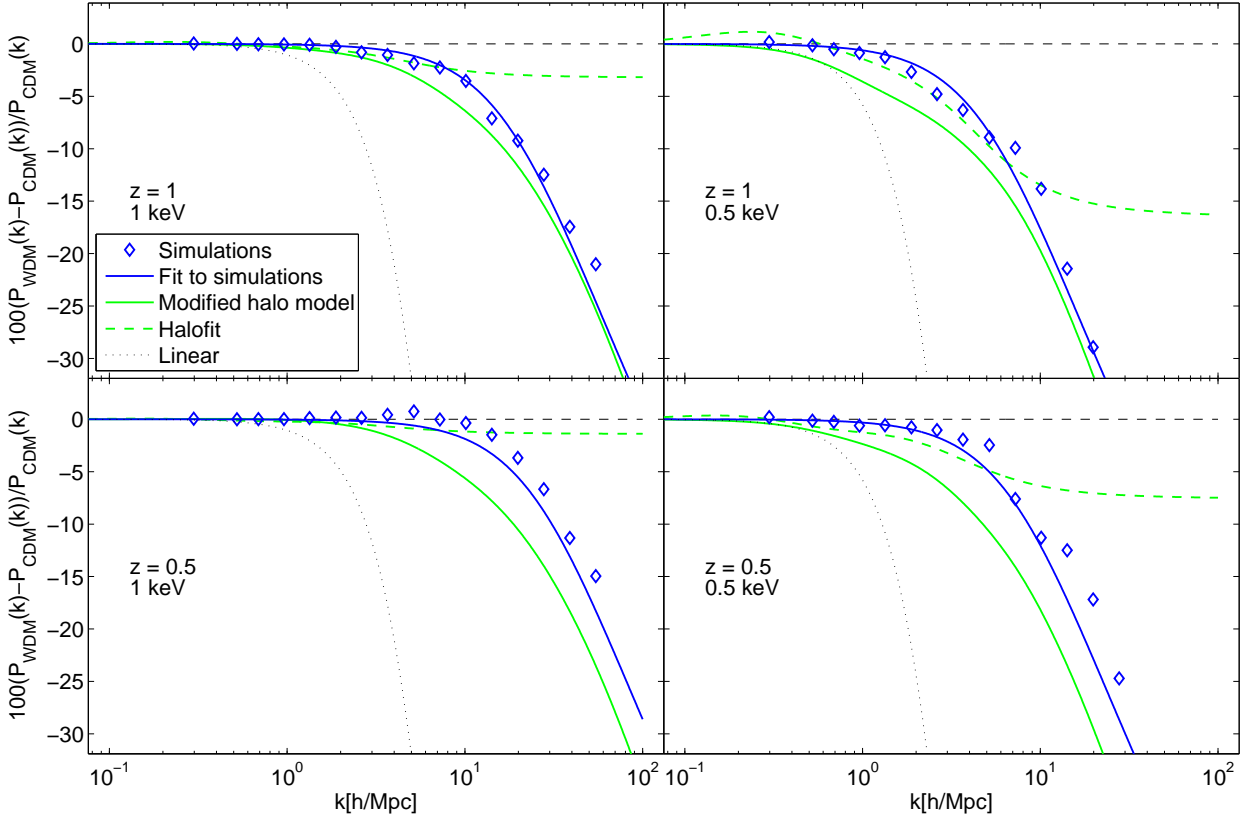}
\end{center}
\caption{The comparison of different non-linear models at redshifts
  1.0 (top panels) and 0.5 (bottom panels) for WDM particles with
  masses 1 keV (left panels) and 0.5 keV (right panels). The blue diamonds represent
  the fractional differences calculated from DM-only simulations from
  previous plots with the fiducial values for $\sigma_{8}$. The blue
  solid lines are the corresponding analytical fits from equation
  \ref{eq_fitting}. The green solid lines are calculated using the
  modified halo model, whereas the green dashed line is the standard
  {\small{HALOFIT}}. The dotted line is the effect as seen in the
  linear matter power spectrum.}
\label{fig_nlmods}
\end{figure*}
\begin{figure*}
\begin{center}
\includegraphics[width=18cm]{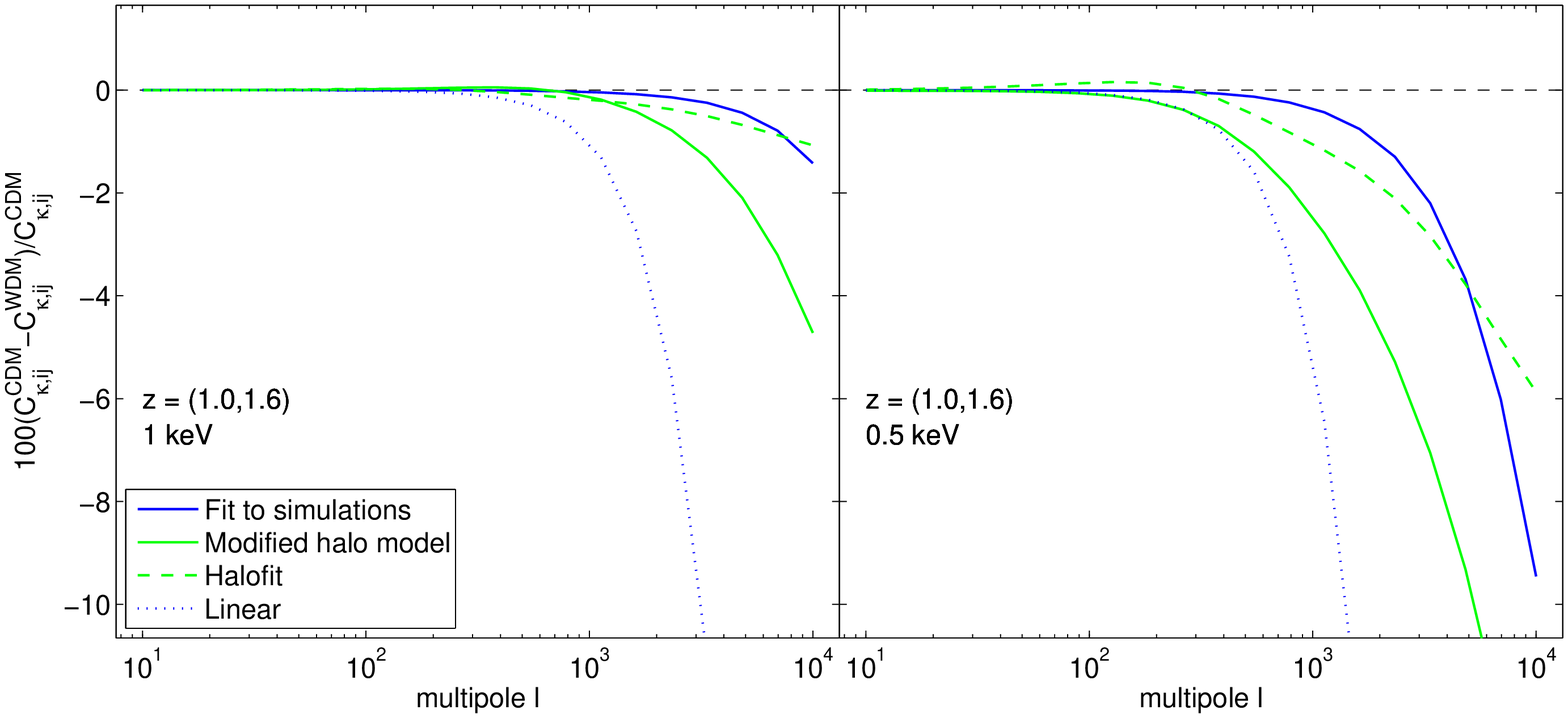}
\end{center}
\caption{The comparison of the impact of using different models of
  non-linear power spectra from figure \ref{fig_nlmods} on the weak
  lensing power spectrum. As above, the blue line is the fractional
  difference in percent between weak lensing power spectra calculated
  using the fitting function found in this work
  (\ref{eq_fitting}). The green solid line is the weak lensing power
  spectrum calculated using the halo model modified for WDM. The
  dashed green line is the same using standard {\small{HALOFIT}}. The
  dotted line is calculated by omitting all non-linear corrections. It
  is evident that excluding such corrections causes a significant
  overestimation of the WDM effect. All the lines in this plot are
  calculated from cross power spectra of the 5th and 8th tomographic
  bins (corresponding to $z=1.6$ and $z=1$, respectively) for WDM
  particle masses of \mwdm=1 keV (left panel) and \mwdm=0.5 keV (right panel).}
\label{fig_Clmods}
\end{figure*}

\section{Comparison with halo model}
As described in Section~\ref{sec_wlps}, it is necessary to have a robust 
model of non-linear structure in order to take full advantage of future weak 
lensing data. For this reason we compare the non-linear matter power spectra
extracted from our simulations with previously derived non-linear models.
The halo model of non-linear structure is based on the assumption that large 
scale structure is made up of individual objects occupying peaks in the matter
overdensity field \citep{press1974,seljak2000,cooray2002}. The most important
elements of this model, the mass function, the halo bias \citep{press1974} and 
the halo density profile \citep{navarro1995} are based on the assumptions that 
all dark matter in the universe is found in haloes and that there is no 
observable suppression of small scale overdensities from early-times 
free-streaming of dark matter particles or late-times thermal velocities. 

These are characteristic properties of CDM, but do not apply to
WDM. For this reason \citet{smith11} modified the halo model by 
applying a specific prescription to the non-linear contribution,
in
addition to suppressing the initial density field, modelled by
applying a transfer function from \citet{viel05} to the linear matter
power spectrum. Such prescription consists of: 
$i)$ treating the dark matter density field as made up
of two components: a smooth, linear component and a non-linear
component, both with power at all scales; $ii)$ introducing a cut-off
mass scale, below which no haloes are found; $iii)$ suppressing the mass
function also above the cut-off scale and $iv)$ suppressing the centers
of halo density profiles by convolving them with a Gaussian function,
whose width depended on the WDM relic thermal velocity. 

Here, we do
not attempt to explore each of these elements with simulations
individually, but rather compare the final matter power spectra found
from simulations with those from the WDM halo model of \citet{smith11}.

Secondly, \cite{smith} compared the standard CDM halo model to CDM simulations
of large scale structure formation and developed an analytical fit to the 
non-linear corrections of the matter power spectrum, known as {\small{HALOFIT}}. 
We apply these corrections to a linear matter power suppressed by the 
\cite{viel05} WDM transfer function (see Equation \ref{eq1}).

We show the results of these comparisons in Figure
\ref{fig_nlmods}. As before, we plot the percent differences between
the WDM and CDM matter power spectra obtained from our simulations of
WDM only. We show this for particle masses of \mwdm=1 keV (left panels) and
\mwdm=0.5 keV (right panels) at redshifts $z=1$ (top row) and $z=0.5$ (bottom row).  We
find that the WDM halo model is closest to simulations at redshift 1
for 1 keV WDM, but that it over-estimates the suppression effect at
redshift 0.5 for 0.5 keV WDM by about 5 percent on scales $k>1$. On
scales $k<1$ \ihmpc however, the {\small{HALOFIT}} non-linear
correction describes the simulations better than the halo model, even
though on smaller scales it severely underestimates the suppression
effect, which becomes worse at lower redshifts. A further small
modification of the WDM halo model will improve its correspondence to
the simulations and allow one to use it at small scales.

We additionally consider these models of non-linear WDM structure to
calculate the weak lensing power spectra in order to explore the
significance of using the correct model. We again plot percentage
differences between WDM and CDM weak lensing power spectra in Figure
\ref{fig_Clmods}. We show only curves representing the cross
correlation power spectrum of redshift bins at $z=1$ and $z=1.6$ for
consistency with Figure \ref{fig_Clbins}. We again examine WDM models
with particle masses of \mwdm=1 keV (left panel) and \mwdm=0.5 keV
(right panel). We also calculate the weak lensing power spectra
without non-linear corrections to the matter power spectrum and note
that this severely over-estimates the effect of WDM suppression. In
the lensing calculation, the {\small{HALOFIT}} non-linear corrections
applied to the WDM suppressed linear matter power spectrum seem to
perform better in describing the results of our WDM simulations than
than the WDM halo model. This due to the fact that the range of
wavenumbers that are better described by the {\small{HALOFIT}}
corrections, namely $k<1$\ihmpc\, are significantly more relevant to
the weak lensing power spectrum than the smaller scales where
{\small{HALOFIT}} strongly deviates from the simulation results.
%

%%%%%%%%%%%%%%%%%%%%%%%%%%%%%%%%%%%%%%%%%%%%%%%%

\section{Conclusions}
By using a large set of N-body and hydrodynamic simulations we have
explored the non-linear evolution of the total matter power.  The
focus of the present work is on small scales and relatively low
redshifts where non-linear effects are important and need to be
properly modelled with simulations.  We checked for numerical
convergence and box-sizes/resolution effects in the range $k=1-10$
\ihmpc. We explored how different masses of a warm dark matter
candidate affect the non-linear suppression as compared to a
corresponding $\Lambda$CDM model that shares the same parameters and
astrophysical inputs.
Our findings can be summarized as follows:
\begin{itemize}
\item[-] Cosmological volumes of linear size 25$h^{-1}$ comoving Mpc
  and with $512^3$ DM particles are sufficient to sample the WDM
  suppression for \mwdm $\ge 1$ keV at the percent level at $k<10$
    \ihmpc.
\item[-]
The non-linear suppression induced by WDM is strongly redshift dependent.
However, by $z=0$, up to $k=10$ \ihmpc, there are virtually no differences
(below 1\%) between $\Lambda$CDM and WDM models with \mwdm $\ge$ 1 keV.
\item[-] At higher redshifts differences are larger, being closer to
  the linear suppression. At $z\sim 1$ there are differences of
  the order of a few percent between the non-linear WDM and
  $\Lambda$CDM power spectra.
\item[-] Baryonic physics and in particular radiative processes in
  the gas component, the star formation criterion and galactic
  feedback in the form of winds are likely to affect the matter power
  at the 2-3 \% level in the range $k=1-10$ \ihmpc. However, a
    much stronger effect can be expected if AGN feedback is
    considered as in (\cite{vandaalen11}).
\item[-] We investigate how a change in $\Omega_{\rm m}$, $H_0$ and
  $\sigma_8$ impacts the non-linear power and
  WDM suppression in particular, when values different from our reference choice are
  used. Small difference are found (at the $\pm$ 2\% level) at the scales
  considered here. Thus the WDM cutoff has a distinctive feature which
  is not degenerate with other cosmological parameters also at a
  non-linear level.
\item[-] We provide a useful fit to the non-linear WDM induced
  suppression in terms of a redshift-dependent transfer function; this
  fitting formula should agree to the actual measured power at the 2\%
  level at $z<3$ and for masses above 0.5 keV. 
\item[-] Reaching a higher accuracy (percent level) in terms of WDM
  non-linear power would require a very extensive analysis of
  astrophysical aspects related to the baryonic component such as
  considering different feedback effects. Among these the most
  promising seems to be AGN feedback which happen to solve the
  overcooling problem and has a strong impact on the total matter
  power (see \cite{vandaalen11,semboloni11}).
\item[-] We find that future weak lensing surveys will most likely be
  powerful enough to rule out WDM masses smaller than 1 keV, which is
  consistent with previous results of \cite{markovic11} and
  \cite{smith11}. Ruling out models for masses larger than 1 keV would
  still be possible by using the cross-correlation signal between
  different redshift bins. However, measurement of the weak
    lensing power at these scales should also consider the effect of
    baryonic physics carefully and parameters could be biased as
    recently found in \cite{semboloni11}, where it has been shown that
    AGN feedback produces a suppression which is larger than the one
    induced by WDM at the scales considered here.
\item[-] 
  Non-linear corrections to the matter power spectrum in the WDM
  scenario obtained from {\small{HALOFIT}} correspond better to the
  results of the WDM only simulation at scales $k<10$ \ihmpc, if
  compared to the non-linear corrections of the halo model from Smith
  \& Markovic (2011). Because these scales are most relevant for weak
  lensing power spectra, using {\small{HALOFIT}} yields a better
  correspondence to the weak lensing power spectra calculated using
  our fitting function. However, on scales $k>10$ \ihmpc, the halo model
   performs slightly better in that it better describes
  the shape of the suppression in the power spectrum, even if it does
  overestimate the effect. For this reason we believe that a further
  modification to the halo model may be needed, especially for weak
  lensing power spectra calculations.
\end{itemize}

  As recently shown
  by \cite{vandaalen11} and \cite{semboloni11} including AGN feedback
  has strong consequences in terms of matter power and weak lensing, a
  comprehensive analysis that aims at measuring the mass of a warm
  dark matter candidate should thus hope to lift the degeneracies
  present (i.e. suppression in terms of matter power) by exploiting
  the different redshift and scale dependencies and by fully exploring
  the astrophysical parameter space and marginalize over the nuisance
  parameters.

We believe that future efforts aiming at measuring the coldness of
cold dark matter should investigate the non-linear matter power in the
range $z=0-5$ either using weak lensing observables or the small scale
clustering of galaxies. These constraints can be particularly useful
since they are complementary to those that can be obtained from high
redshift \lya forest data (e.g. BOSS/SDSS-III survey) or galactic and
sub-galactic observables in the local universe.

\section*{Acknowledgments.}
We are grateful to Robert Smith for providing tables to calculate the
halo model power spectrum and to the referee Joop Schaye for his comments.
Numerical computations were performed at the High Performance Computer
Cluster Darwin (HPCF) in Cambridge (UK) and at the RZG computing
center in Garching.  We acknowledge support from an ASI/AAE grant, ASI
contracts Euclid-IC (I/031/10/0), INFN PD51, PRIN MIUR, PRIN INAF
2009, ERC-StG ``cosmoIGM'', the DFG Cluster of Excellence ``Origin
and Structure of the Universe'', the TRR33 Transregio Collaborative Research
Network on the ``Dark Universe'', and from the European Commissions FP7
Marie Curie Initial Training Network CosmoComp (PITN-GA-2009-238356).
Post-processing of the simulations has also been carried at CINECA
(Italy) and COSMOS Supercomputer in Cambridge (UK).  KM acknowledges
support from the International Max-Planck Research School.

\bibliographystyle{mn2e} \bibliography{master2.bib}

\end{document}